\documentclass[pre,superscriptaddress,unsortedaddress,showpacs,amsmath,amssymb,endfloats, twocolumn]{revtex4}
\def\figwidth{0.9\linewidth}
\usepackage[final]{graphicx}

\begin{document}
\title{Aging in short-ranged attractive colloids: A numerical study}
\date{\today}
\def\roma{
  \affiliation{Dipartimento di Fisica and INFM Center for Statistical Mechanics  and Complexity, Universit\`a di Roma ``La Sapienza'', Piazzale  Aldo Moro 2, 00185 Roma, Italy}}
\def\boston{
  \affiliation{ Center for Polymer Studies and
  Department of Physics, Boston University, Boston, MA 02215, USA.}}
\author{G.~Foffi}\roma
\author{E. Zaccarelli}\roma
\author{S. Buldyrev}\boston
\author{F.~Sciortino}\roma
\author{P.~Tartaglia}\roma

\begin{abstract}
We study the aging dynamics in a model for dense simple liquids, in
which particles interact through a hard-core repulsion complemented by
a short-ranged attractive potential, of the kind found in colloidal
suspensions.  In this system, at large packing fractions, kinetically
arrested disordered states can be created both on cooling (attractive
glass) and on heating (repulsive glass).  The possibility of having
two distinct glasses, at the same packing fraction, with two different
dynamics offers the unique possibility of comparing --- within the
same model --- the differences in aging dynamics. We find that, while
the aging dynamics of the repulsive glass is similar to the one
observed in atomic and molecular systems, the aging dynamics of the
attractive glass shows novel unexpected features.

\end{abstract}

\maketitle

\section{Introduction}
In the last years, the physics of systems characterized by
short-ranged attractive inter-particle interactions --- short in
comparison to the particle size --- has been the focus of several
investigations. A good example of this type of systems are colloidal
solutions, since the intensity and the range of the potential can be
finely controlled by changing the solvent properties or the chemistry
of the dispersed particles.  The short range of the attractive
interaction produces thermodynamic features which are not observed in
atomic or molecular system (where the interaction range is always long
ranged)\cite{Pusey1991,Gast83,Meijer1991,Anderson2002}.  Recently,
theoretical, numerical and experimental studies have focused on the
dynamical properties of these systems. One of the most astonishing
discovery is that, at high density, the metastable liquid is
characterized by a non monotonic temperature dependence of the
characteristic structural times: the dynamics slows down not only upon
cooling (as commonly observed in molecular systems), but also upon
heating. The slowing down upon heating can be so intense that a novel
mechanism of arrest takes place high $T$ \cite{Sciortino2002b}.  In
the high density part of the phase diagram, for sufficiently
short-ranged potentials, a re-entrant liquid-glass line is observed
and a liquid phase emerges between two glasses. In other words, moving
along a constant density path, it is possible to pass from a glass
phase to a liquid phase and then again in a new glass phase just by
progressively lowering the temperature. The two glasses are named
``attractive glass'' and ``repulsive glass'' due to the different
mechanisms, respectively attraction and excluded volume, which control
the dynamical slowing down \cite{NOTA}.  In the re-entrant liquid
region, between the attractive and repulsive glasses, dynamics is very
unusual. Correlation functions show a subtle logarithmic decay, while
the particle mean squared displacement follows a power-law in time.

The complex dynamics in short-ranged attractive systems has been
interpreted within Mode Coupling Theory (MCT)
\cite{Goetze1991b}. Theoretical predictions do not depend on the
detailed shape of the attractive potential and have been confirmed for
many different modelization of the attractive tail.
\cite{Fabbian1999,Bergenholtz1999,Dawson2001,Foffi2002,Sperl2003}. 
A set of
simulations\cite{Puertas2002,Foffi2002c,Zaccarelli2002b,Puertas2003}
and experiments
\cite{Mallamace2000,Pham2002,Eckert2002,Chen2003,Chen2003b,Bartsch2003}
have confirmed the existence of such complex dynamics.\\ 
In numerical work based on the Square Well System (SWS), one of the
simplest models studied, it has been shown that a diffusivity maximum
in temperature appears when the width of the square well is short enough
compared to the hard core diameter \cite{Zaccarelli2002b}.
The calculation of the iso-diffusivity lines in equilibrium
\cite{Foffi2002c} confirms
the reentrant behavior and the novel dynamics which take place in the
reentrant region. Along an iso-diffusivity path, the decay of the
density-density correlation functions, which at high temperature can
be well described by a stretched exponential, becomes more and more
uncommon upon cooling, showing at low T signs of logarithmic
decay\cite{Puertas2002,Zaccarelli2002b,Sciortino2003pre}.\\ Previous
numerical work has focused on equilibrium properties and on the
dynamics near the predicted glass-glass
transition~\cite{Puertas2002,Foffi2002c,Zaccarelli2002b,Zaccarelli2003b}. Here
we report a study of the non-equilibrium dynamics. Starting from an
equilibrium state in the re-entrant fluid region, the system is
quenched both at low and high temperature.
We are then able to explore --- within the same model --- the aging
dynamics for the attractive and repulsive glass, to find out if the
differences which characterize the dynamics in the glassy state show
up also in the out-of-equilibrium evolution. We discover that, while
the aging dynamics of the repulsive glass is similar to the one
observed in atomic and molecular
liquids\cite{Kob1997b,Sciortino2001a}, the aging dynamics of the
attractive glass shows novel unexpected features. We also report
comparison with a recent experimental study of aging in short ranged
colloidal systems \cite{Pham2003pre}
\begin{figure}
\includegraphics*[width=\figwidth]{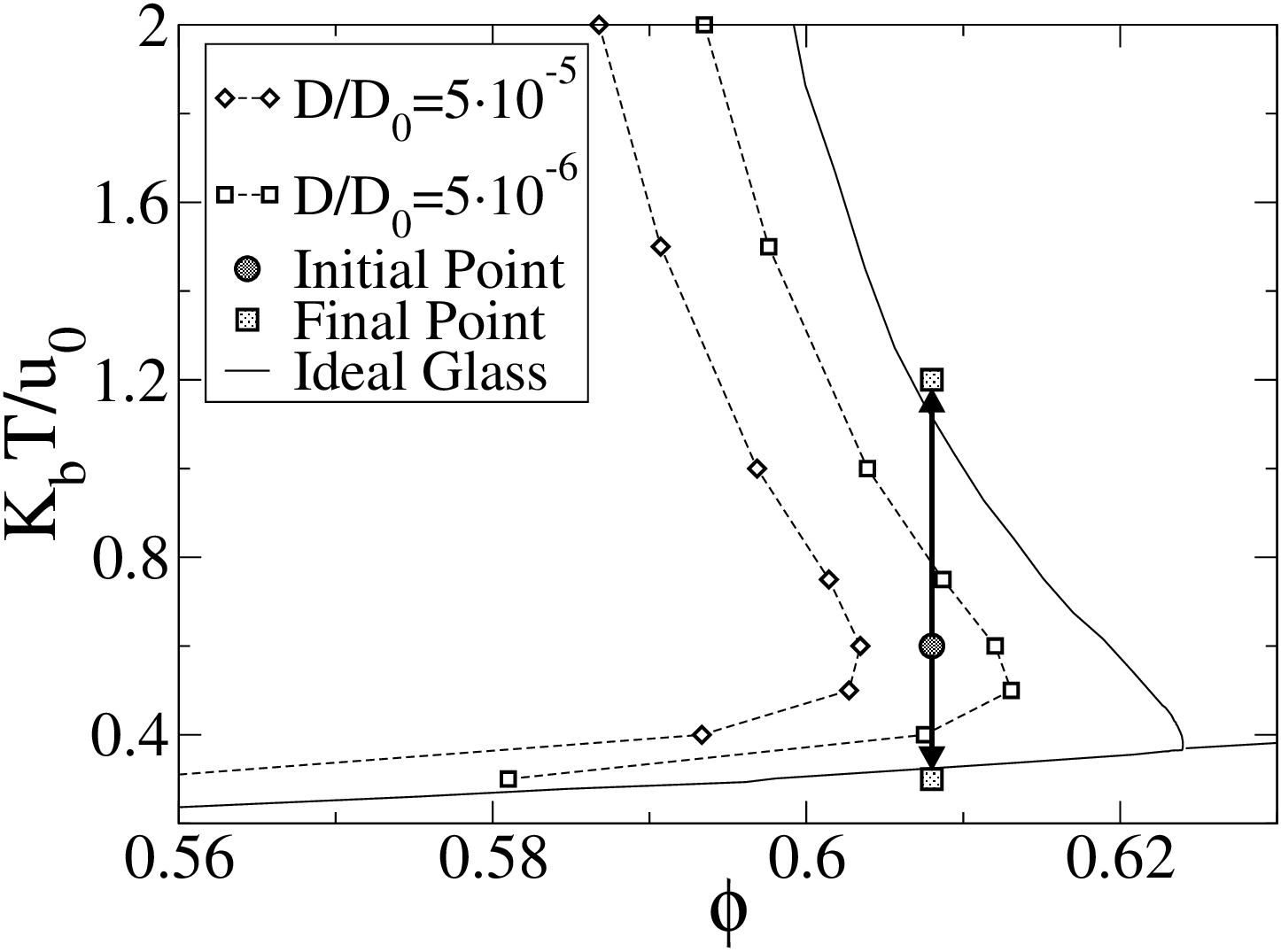}
\caption{\label{fig1}
	Square Well iso-diffusivity curves (i.e. the locus of points in
	the $\phi,T$ plane where the normalized diffusion coefficient
	is constant) for three different values of $D/D_0$
	($D_0=\sigma_b \sqrt{T/m}$ data from
	Ref.\protect\cite{Zaccarelli2002}).  The extrapolated ideal
	glass line is also reported (data from
	Ref.\protect\cite{Sciortino2003pre}). Equilibrium starting
	configurations at $\phi=0.608$ and $T_i=0.6$ (filled circle)
	are instantaneously quenched at $t=0$ to $T_f=0.3$ and
	$T_f=1.2$ (plus).}
\end{figure}

\section{Simulation details}

We investigate a system that has been extensively studied earlier in
the equilibrium regime, a binary square well
mixture\cite{Zaccarelli2002b}.  The binary system is a $50\%$-$50\%$
mixture of $N=700$ particles. The two species (labeled $A$ and $B$)
are characterized by a diameter ratio
$\delta=\sigma_A/\sigma_B=1.2$. Masses are chosen to be equal and
unitary. The attraction is modeled by a square well interaction
defined according to:
\begin{equation}
V_{ij}(r) = \left\{
  \begin{array}{ll}
    \infty & \mbox{ $r_{ij}<\sigma_{ij}$}
\\ 
    -u_0 & \mbox{$\sigma_{ij}<r_{ij}<\sigma_{ij}+\Delta_{ij}$}
\\
    0 & \mbox{$r_{ij}>\sigma_{ij}+\Delta_{ij}$}
  \end{array} \right.
\end{equation}
where $\sigma_{ij}=(\sigma_i+\sigma_j)/2$, $i,j=A,B$ and $\Delta_{ij}$
has been chosen in such a way that
$\epsilon_{ij}=\Delta_{ij}/(d_{ij}+\Delta_{ij})=0.03$. We choose
$k_B=1$ and the depth of the potential $u_0=1$. Hence $T=1$
corresponds to a thermal energy equal to the attractive well
depth. The diameter of the small specie is chosen as unity of length,
i.e. $\sigma_B=1$. Density is parametrized in terms of packing
fraction $\phi=(\rho_A d_A^3+\rho_B d_B^3)\cdot \pi/6$, where
$\rho_i=N_i/L^3$, $L$ being the box size and $N_i$ the number of
particles for each species. Time is measured in units of
$\sigma_B\cdot(m/u_0)^{1/2}$. The unit has been chosen consistently
with Ref.~\cite{Zaccarelli2002b}, where the same system has been
studied in equilibrium. A standard MD algorithm has been implemented
for particles interacting with SWS potentials
\cite{Rapaport97}. Between collisions, particles move along straight
lines with constant velocities. When the distance between the
particles becomes equal to the distance where $V(r)$ has a
discontinuity, the velocities of the interacting particles
instantaneously change.  The algorithm calculates the shortest
collision time in the system and propagate the trajectory from one
collision to the next one.  Calculations of the next collision time
are optimized by dividing the system in small subsystems, so that
collision times are computed only between particles in the neighboring
subsystems.\\
With the present choice of parameters, the formation of crystallites
during the simulation run is not observed.  As compared to the
monodisperse case, the slight asymmetry in the diameters allows us to
follow the development of the slow dynamics over several orders of
magnitude\cite{Foffi2002c}. We choose to work at the packing fraction
$\phi=0.608$, being this density in the middle of the re-entrant
region of the phase diagram, as shown in previous calculations
\cite{Zaccarelli2002b}. At this density we prepared 60 independent
configurations, equilibrated at the initial temperature
$T_i=0.6$. Subsequently we quenched these independent configurations
to the desired final temperatures, and then we followed the evolution
in time at constant temperature.  The characteristic time of the
thermostat as been chosen to be much smaller than the structural
relaxation time and such that the system may equilibrate within one
time unit.  Each of the 60 independent configurations has been
quenched to $T_f=1.2$, i.e. in the repulsive glass, and to $T_f=0.3$,
i.e. in the attractive glass and run up to $t_f=2
\cdot10^5$. The numerical protocol is illustrated in
Fig.~\ref{fig1}. 
Iso-diffusivity curves from Ref.~\cite{Zaccarelli2002b} and the
extrapolated ideal glass line from Ref.~\cite{Sciortino2003pre} are
plotted together with the quench path selected in this work.
Since the system, after the quench, is out of equilibrium, all
averages --- represented in the text by the symbol $\langle\cdot
\rangle$ --- have been performed over the ensemble of the 60 initial
independent configurations.

\section{Results and Discussion}
\begin{figure}
\includegraphics*[width=\figwidth]{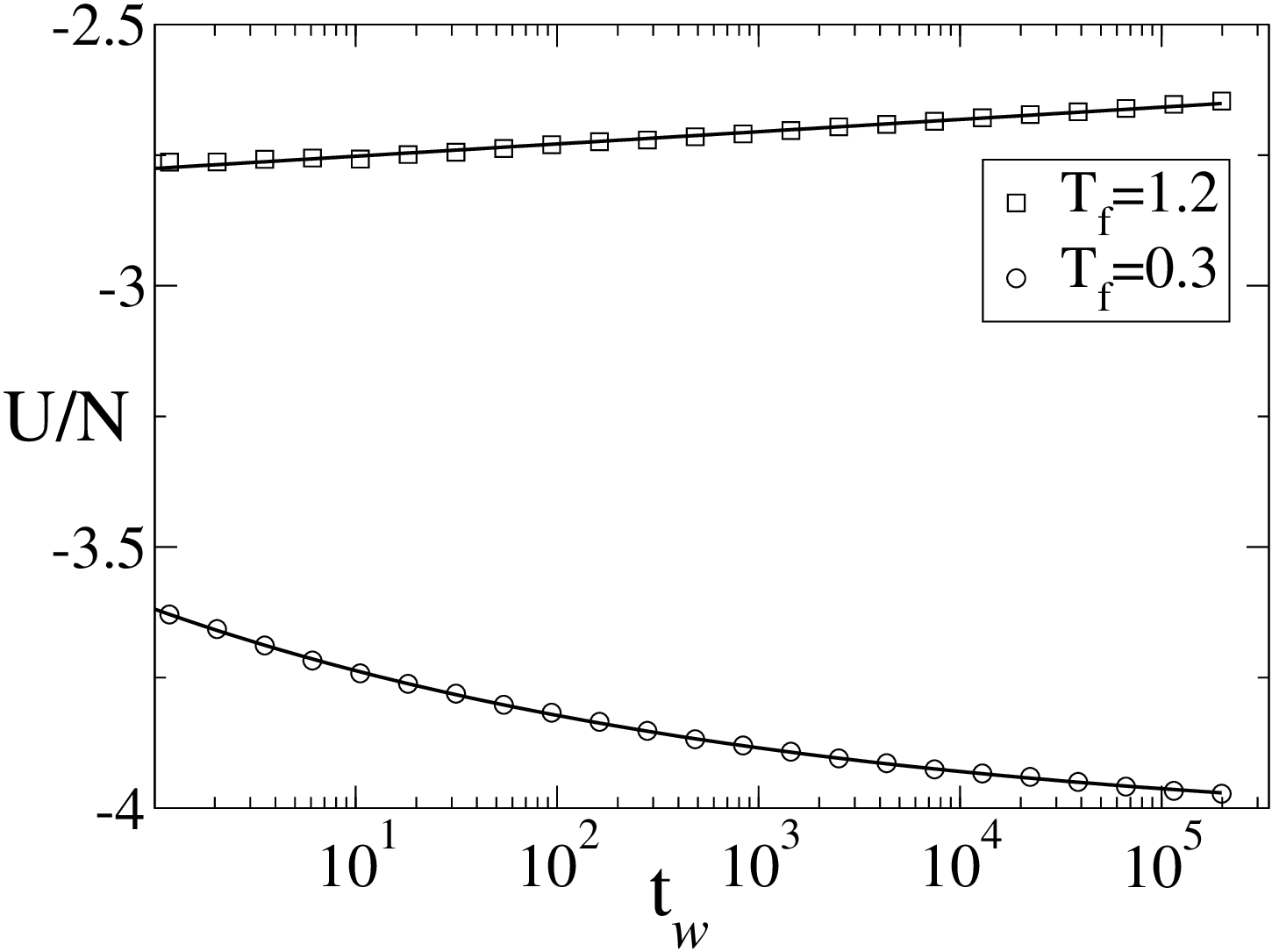}
\caption{\label{fig2}  
	Average potential energy $U$ after the quench as function of
	the time $t_w$ elapsed from the quench. The symbols are the
	simulation data and the continuous line are the fits. Data for
	$t_w<1$ are not shown since the kinetic energy has mot yet
	equilibrated to the final value. For $T_f=1.2$ the fit is
	$U/N=-2.77 +0.01\ln{(t_w)}$ while for $T_f=0.3$ we obtain
	$U/N=-4.05 +0.43\,t_w^{-0.14}$}
\end{figure}

\begin{figure}
\includegraphics*[width=\figwidth]{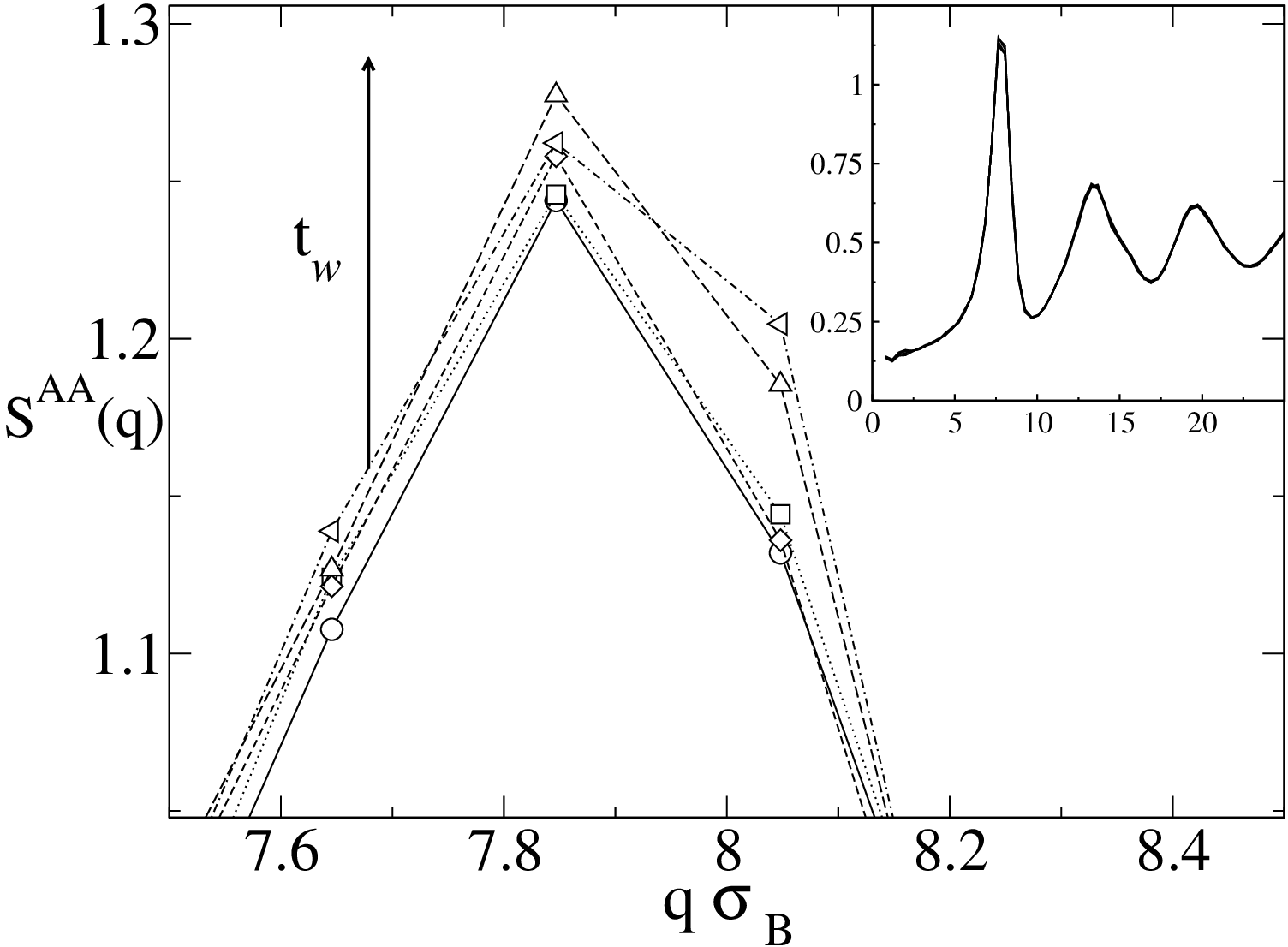}
\caption{\label{fig3} 
          First peak of the static structure factor at different
          waiting time $t_w$, $10^1$ (circles), $10^2$ (squares),
          $10^3$ (diamonds), $10^4$ (triangles up) and $10^5$(triangles
          left). The quench is at $T_f=1.2$. For long $t_w$ there is
          slow growth of the peaks of the structure factors. In the
          inset  structure factors for the same $t_w$  are displayed.}
\end{figure}

Numerical investigations of aging in glassy materials are usually
accomplished by rapidly quenching the system from an equilibrium state
to a new state, where the characteristic time scale of the structural
relaxation is much longer than the observation time. When the
time scale of the final state exceeds the characteristic time of the
simulation, equilibrium cannot be reached within the simulation time.
Out of equilibrium time translation invariance does not hold and the
state of the system does not only depend from the external parameters,
but also
from the time elapsed from the quench, the so-called waiting time
$t_w$. Indeed in the aging regime other notable
effects emerge, i.e. violation of the fluctuation dissipation theorem
\cite{Cugliandolo1993,Parisi1997,Latz2000,Sciortino2001aa,DeGregorio2002} and
strong dependence on the story of the system \cite{Kob1997b}. Thus, in
out-of-equilibrium conditions correlation functions will depend both
on the time of observation $t$ and on the waiting time $t_w$,
i.e. $C_A(t_w, t)=\langle A^*(t_w)A(t_w+t)\rangle$. Similarly, the one
time properties (e.g. the static structure factor or the energy), will
depend on the waiting time $t_w$.\\
The first quantity that we investigate is the average potential energy
of the system $U$.  At $t_w=0^+$, i.e. instantaneously after the
quenches to respectively $T_f=1.2$ and $T_f=0.3$, the two systems
possess the same potential energy. Then, with $t_w$, they start
evolving toward different energy values. The situation is represented
in Fig.~\ref{fig2}.  While the system quenched to $T_f=1.2$, i.e. in
the repulsive regime, shows an increase in the potential energy
(i.e. a progressive breaking of the interparticle bonding), the one
quenched to $T_f=0.3$ shows the opposite behavior.  In both cases, the
time dependence of the potential energy shows a fast short-time ($t_w
< 1$) evolution (which reflects the equilibration of the fast degrees
of freedom) followed by a much slower evolution, proper of the aging
dynamics. The aging dynamics has detectable different trends at high
and low temperatures. While at high temperatures a clear $log(t_w)$
dependence of $U$ is observed, at low $T$ the time dependence in $log$
scale is not linear and can be well fitted by $U-U_{eq} \sim t_w^{-a}$,
with $a \approx 0.14$. A similar form was successfully adopted by
Parisi for a binary system of soft spheres in the aging regime
\cite{Parisi1997}.\\
The time evolution of the aging structure can be followed
by numerical calculation of the partial static structure factors,
$S_{\alpha\beta}(q,t_w)$ defined as
\begin{equation}
S_{\alpha\beta}(q,t_w)=\langle\varrho^*_\alpha({\bf q},t_w)\varrho_\beta({\bf q},t_w)/\sqrt{n_\alpha n_\beta}
\rangle
\end{equation}
where the partial density variables are defined as
$\varrho_\alpha(\vec q)=\sum_k\exp[i {\bf q}\cdot {\bf
r}(t)_k^{(\alpha)}]/\sqrt{N}$, and $n_i=N_i/N$ are the partial
concentrations.  In equilibrium, time translation invariance implies
$S_{\alpha\beta}(q,t)=S_{\alpha\beta}(q,0)$, i.e. the static structure
factor is a time independent function.\\
The inset of Fig.~\ref{fig3} shows $S_{AA}(q,t_w)$ for the quench to
$T_f=1.2$, for waiting times spanning $4$ decades. Similarly to binary
Lennard-Jones systems \cite{Kob1997b}, the variation of the static
structure factor is particularly small, for all $\alpha,\beta$
components.  With increasing $t_w$ a very weak increase
at all peaks of the structure factors is observed. This is shown in
the main figure.  Similarly for the $T_f=0.3$ case, the
$S_{\alpha\beta}(q,t_w)$ changes are very small. Interestingly enough,
for the first peak (and only there) the trend is reverted with respect
to the previous case, i.e. the intensity of $S_{AA}(q,t_w)$ around the
peak decreases on aging. This observation agrees with equilibrium
studies. Indeed, for the square well system, an increase in
temperature produces an increase in the first peak toward the hard
sphere limit whereas lowering the temperature only promotes the
structure factor oscillations at large
wave-vector\cite{Zaccarelli2003d}. It is perhaps interesting to recall
that within MCT, the increase of $S_{\alpha\beta}$ at the first peak
is responsible for the formation of the repulsive glass, whereas the
long lasting oscillations in the structure factor generates the
kinetic arrest in the attractive glass
\cite{Dawson2001}.


\begin{figure}
\includegraphics*[width=\figwidth]{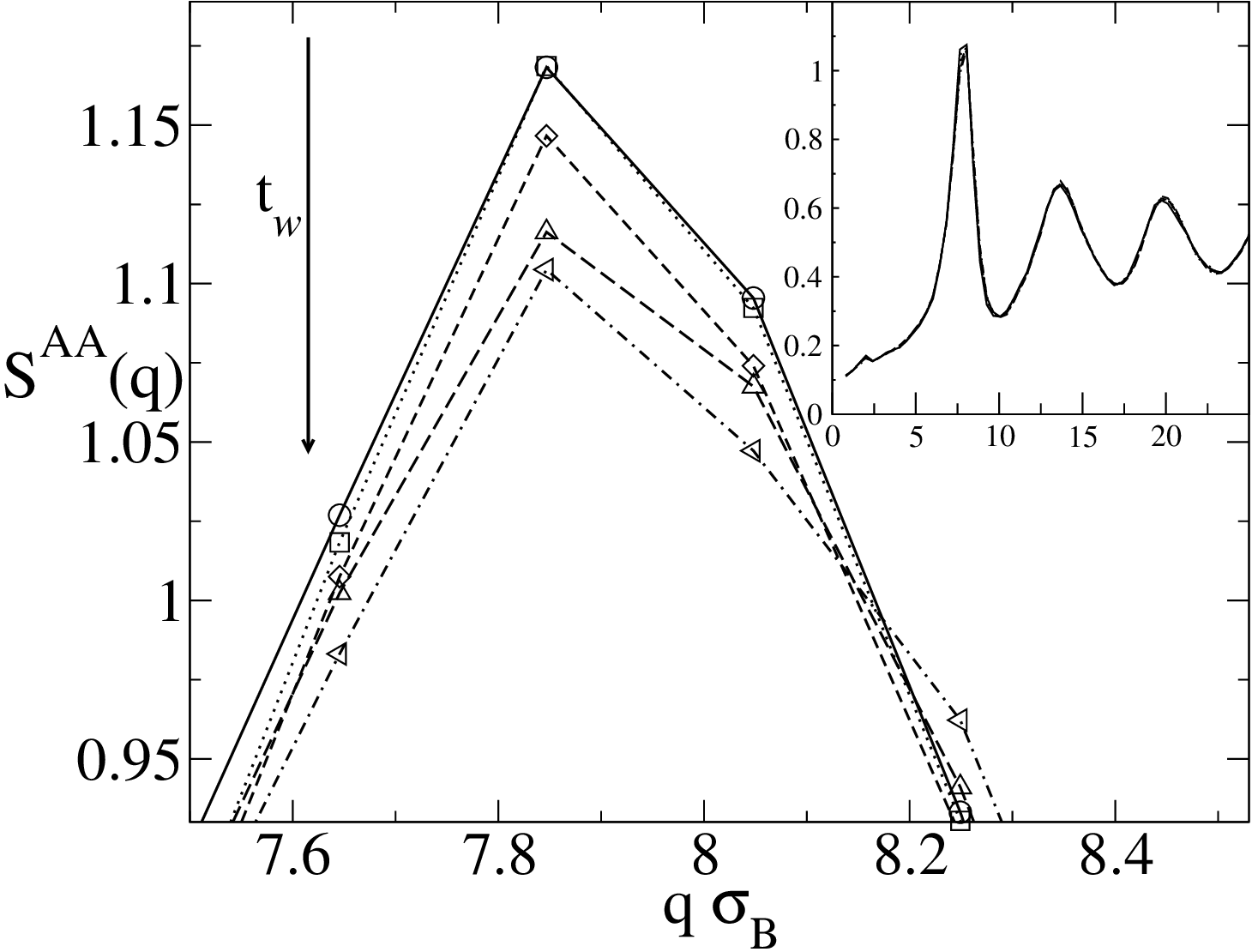}
\caption{\label{fig4} 
	Same as Fig.~\protect\ref{fig3} for $T_f=0.3$.}
\end{figure}

\begin{figure}
\includegraphics*[width=\figwidth]{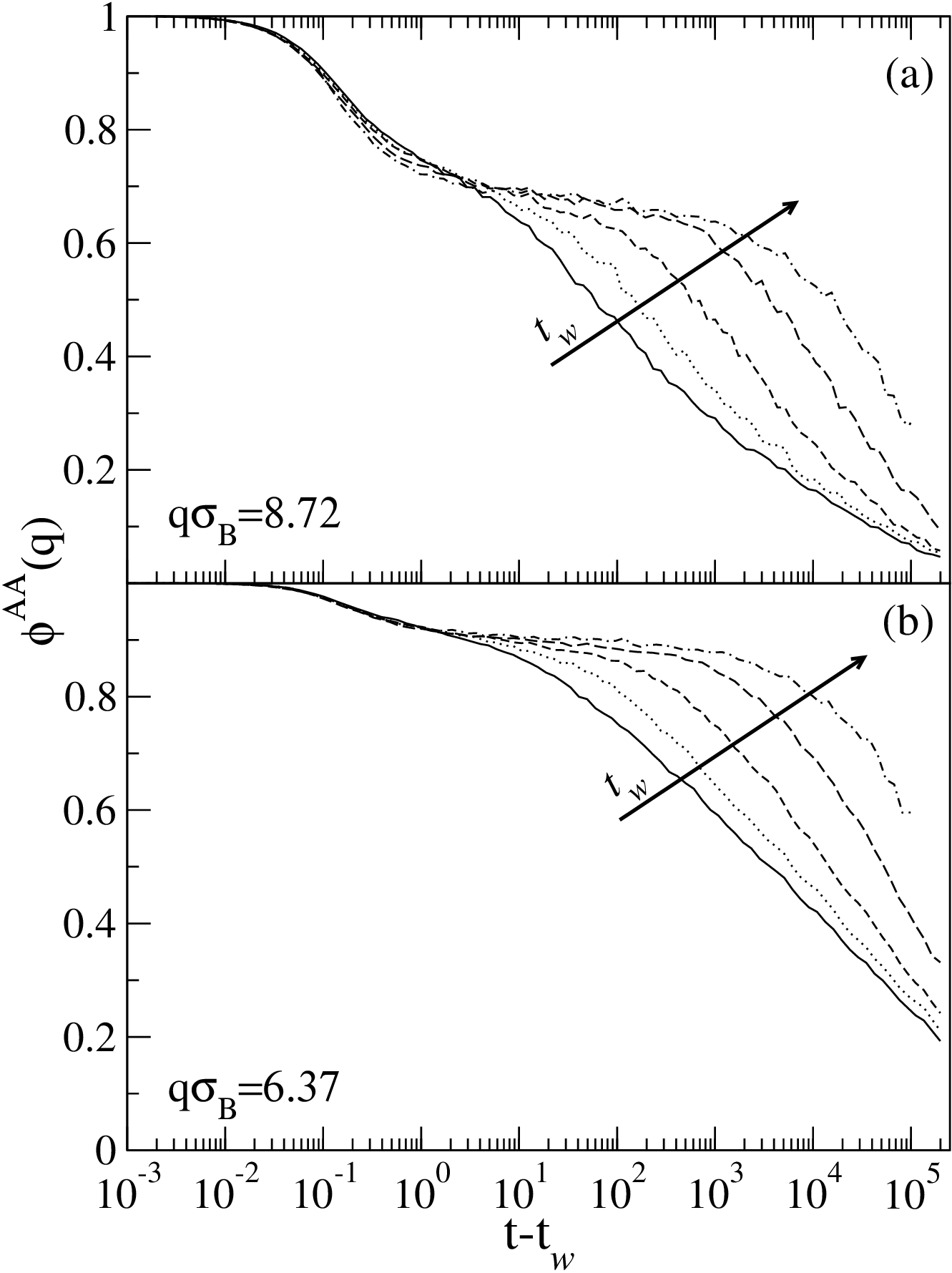}
\caption{\label{fig5} 
    	(a) Density-density correlation function
    	$\phi_{AA}(q,t_w,t-t_w)$ after a quench at $T_f=1.2$
    	for $q\sigma_B=8.72$. From right to left, longer and longer
    	relaxation time, $t_w=10^1$ (continuous line), $10^2$ (dotted
    	line), $10^3$ (short dashed line), $10^4$ (long dashed line),
    	$10^5$ (dot dashed line).  \\(b): same results but for
    	$q\sigma_B=6.37$}
\end{figure}

\begin{figure}
\includegraphics*[width=\figwidth]{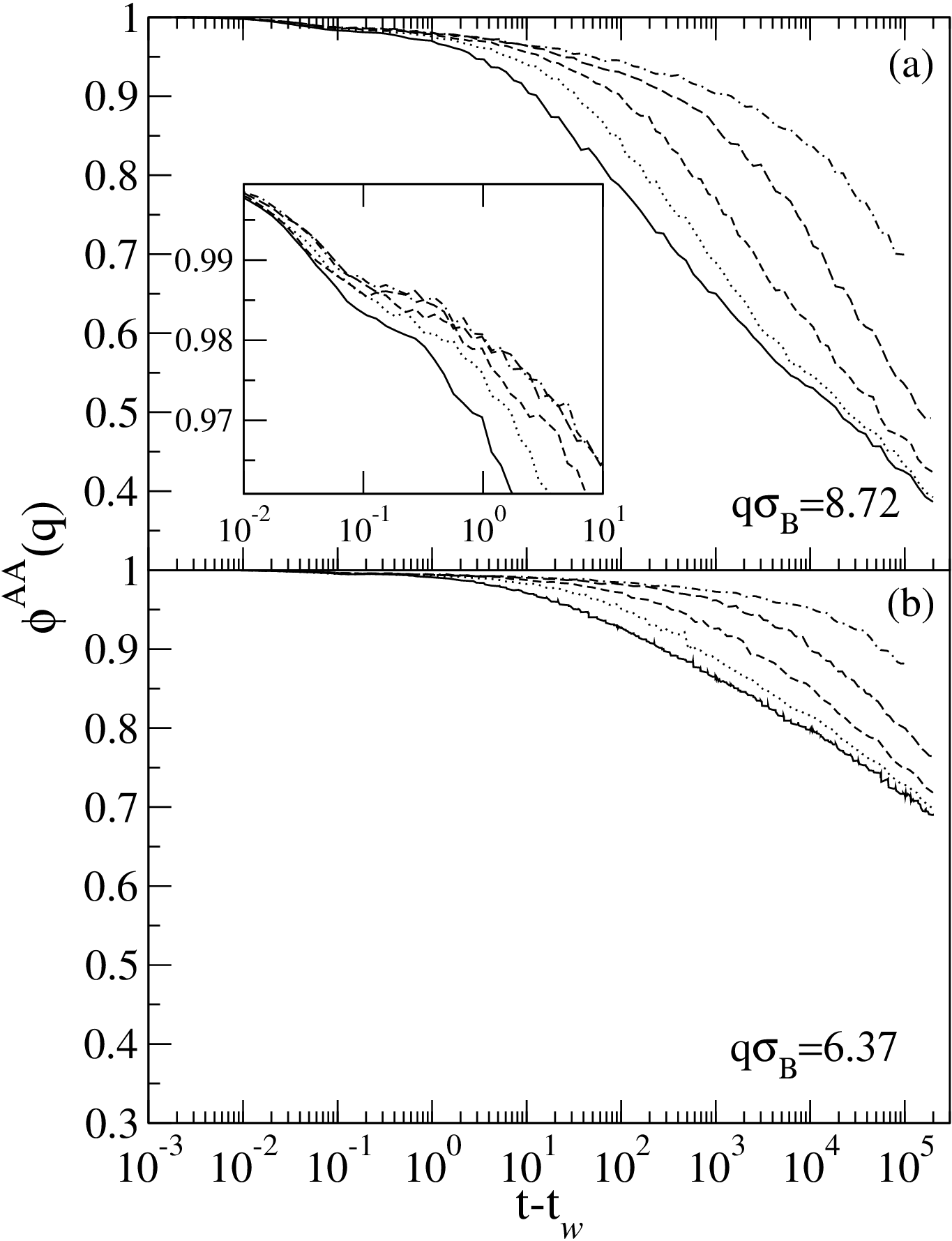}
\caption{\label{fig6} 
  	Same as Fig.~\protect\ref{fig5} but for the quench at
  	$T_f=0.3$. In the inset in (a), an enlargement of
  	the plateau area is shown (see text for details). }
\end{figure}

Next we study the intermediate scattering function, i.e. dynamic
density-density correlators. For a binary mixture this quantity is defined by:
\begin{equation}
\label{fqt}
\phi_{\alpha\beta}(q,t,t')=\langle\varrho_\alpha({\bf q},t)^*\varrho_\beta({\bf q},t')\rangle/S_{\alpha\beta}(q,t')
\end{equation}
where the density variables $\varrho_\alpha({\bf q},t)$ have been
defined before. In equilibrium $\phi_{\alpha\beta}(q,t,t')$ will
depend only on the difference $t-t'$. When the system ages the
density-density correlators is a function of $t_w$ and $t$.  In this
paper, we show $\phi_{\alpha\beta}(q,t,t')$ as a function of $t_w$ and
the time difference $(t-t_w)$,
i.e. $\phi_{\alpha\beta}(q,t_w,t-t_w)$. The density correlation is
normalized by $S_{\alpha\beta}(q,t_w)$.
The upper panel of Fig.~\ref{fig5} shows the $t_w$ dependence of
$\phi_{AA}(q,t_w,t-t_w)$, with $q\sigma_B=8.72$, which lies between
the first and the second peak in the static structure factor. 
On increasing $t_w$, the decay of the correlation functions becomes
slower and slower. The time needed to lose memory of the starting
configuration grows.  This is a typical feature of aging systems and
it has been encountered both in experiments
\cite{Pham2003pre} and simulations \cite{Kob1997b}.
 The slowing down of the dynamics on increasing $t_w$
resembles the way the structural arrest emerges in supercooled liquids
on approaching the glass transition.  Indeed, in this case, upon
cooling two characteristic time scales emerge, a fast decay around a
plateau value and a slow decay to zero.  The time duration of the
plateau becomes longer, the closer the system is to the glass
transition.  On the other hand, in the aging regime
the control parameter is the waiting time $t_w$, which plays a role
analogous to that played by $T$ in equilibrium, i.e. the longer $t_w$
the longer is the time requested to lose memory.  In the same way as
temperature measures the distance from the glass transition in
equilibrium liquids, $t_w$ measures the progressive thermalization of
the structural degrees of freedom. In summary, with increasing $t_w$
the system develops a clear plateau, whose height is consistent with
the value expected for the repulsive glass.\\ Similar results are
shown in the lower panel of Fig.~\ref{fig5} for $q\sigma_B=6.37$,
which is close to the maximum of the static structure factor. Two main
differences emerge between the correlators at the the two wavevectors:
(i) the relaxation time is longer for $q\sigma_B=6.37$, (ii) the
plateau is higher.  Both this issues follow, as expected, the trend of
the equilibrium results, where at the maximum of the structure factor
a maximum of the relaxation time and of the non-ergodicity parameter
is encountered \cite{Foffi2003pre}.\\
The upper panel of Fig.~\ref{fig6} shows the density-density
correlation functions for the quench at $T_f=0.3$ for
$q\sigma_B=8.72$. Similarly to what we found in the previous case, the
time of decay of the correlators grows with the waiting time
$t_w$. However the shape of the functions is drastically different,
confirming that even the aging dynamics reflects the differences noted
in the previous equilibrium studies. Two notable differences with the
$T_f=1.2$ case are: (i) the strength of the
$\alpha$-relaxation(i.e. the height of the plateau) in aging is much
more intense, being now close to one ; (ii) the plateau does not
significantly stretch in time (as shown in the inset) beyond
$t-t_w\approx 1$.  Similar trends are found for a $q$-vector closer to the
first peak of the structure factor, as shown in the lower panel of
Fig.~\ref{fig6}.




A recent study of the glass-glass transition has suggested the
possibility that the attractive glass is significantly destabilized by
hopping processes \cite{Zaccarelli2003b}, which in the present context
can be associated with the breaking of the interparticle bonds.  The
data shown in the inset of Fig.\ref{fig6} support such interpretation,
since no clear plateau can be observed for $t_w> 1$.


%
%

\begin{figure}
\includegraphics*[width=\figwidth]{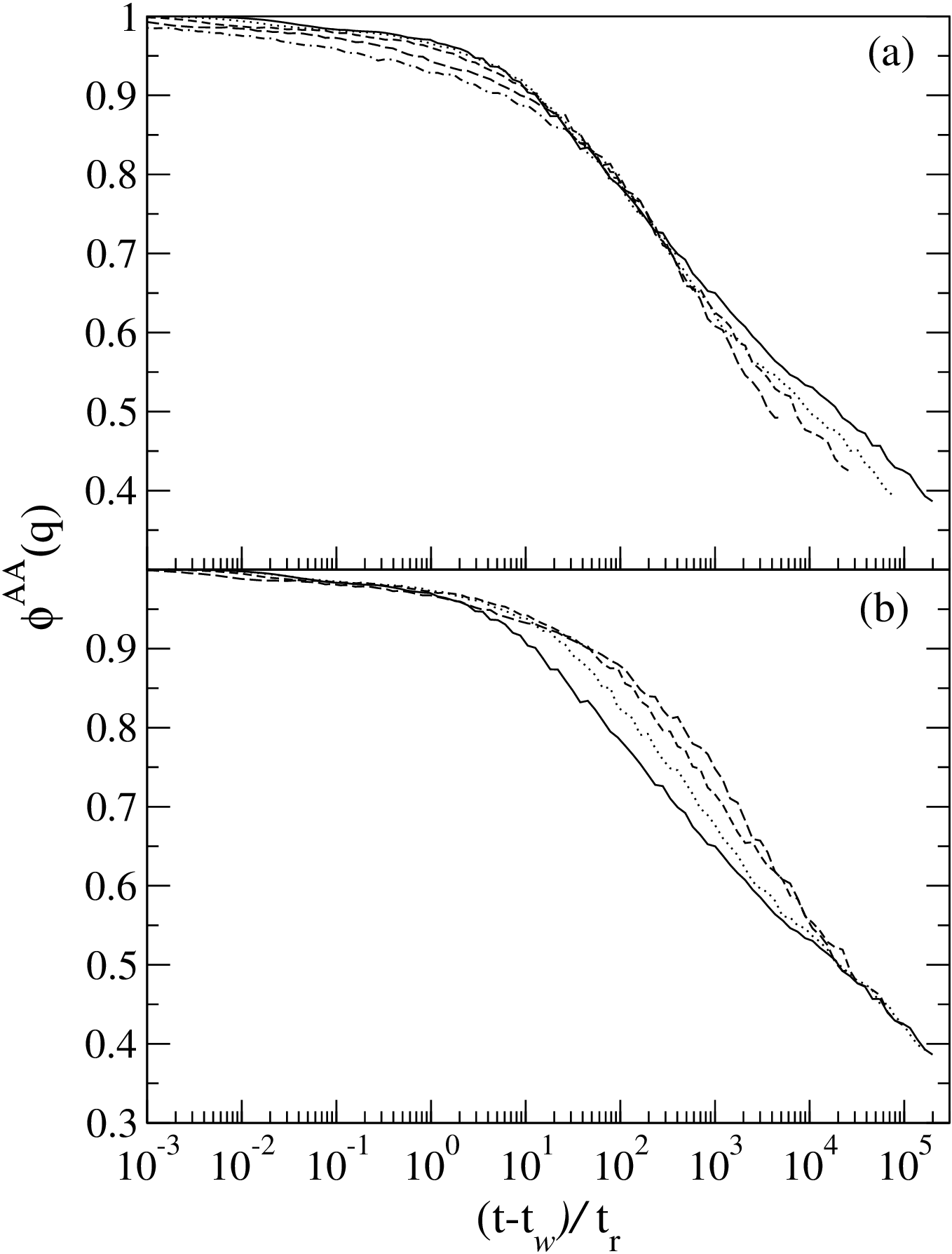}
\caption{\label{fig7} 
	Scaling of the density-density correlation functions for
	$T_f=0.3$. The waiting times are $t_w=10^1$ (continuous line),
	$10^2$ (dotted line), $10^3$ (short dashed line), $10^4$ (long
	dashed line), $10^5$ (dot dashed line). \\(a) best scaling for
	the region where $0.7<\phi_q<0.9$.  \\ (b) best scaling for
	the region where $0.35<\phi_q<0.5$.}
\end{figure}
\begin{figure}
\includegraphics*[width=\figwidth]{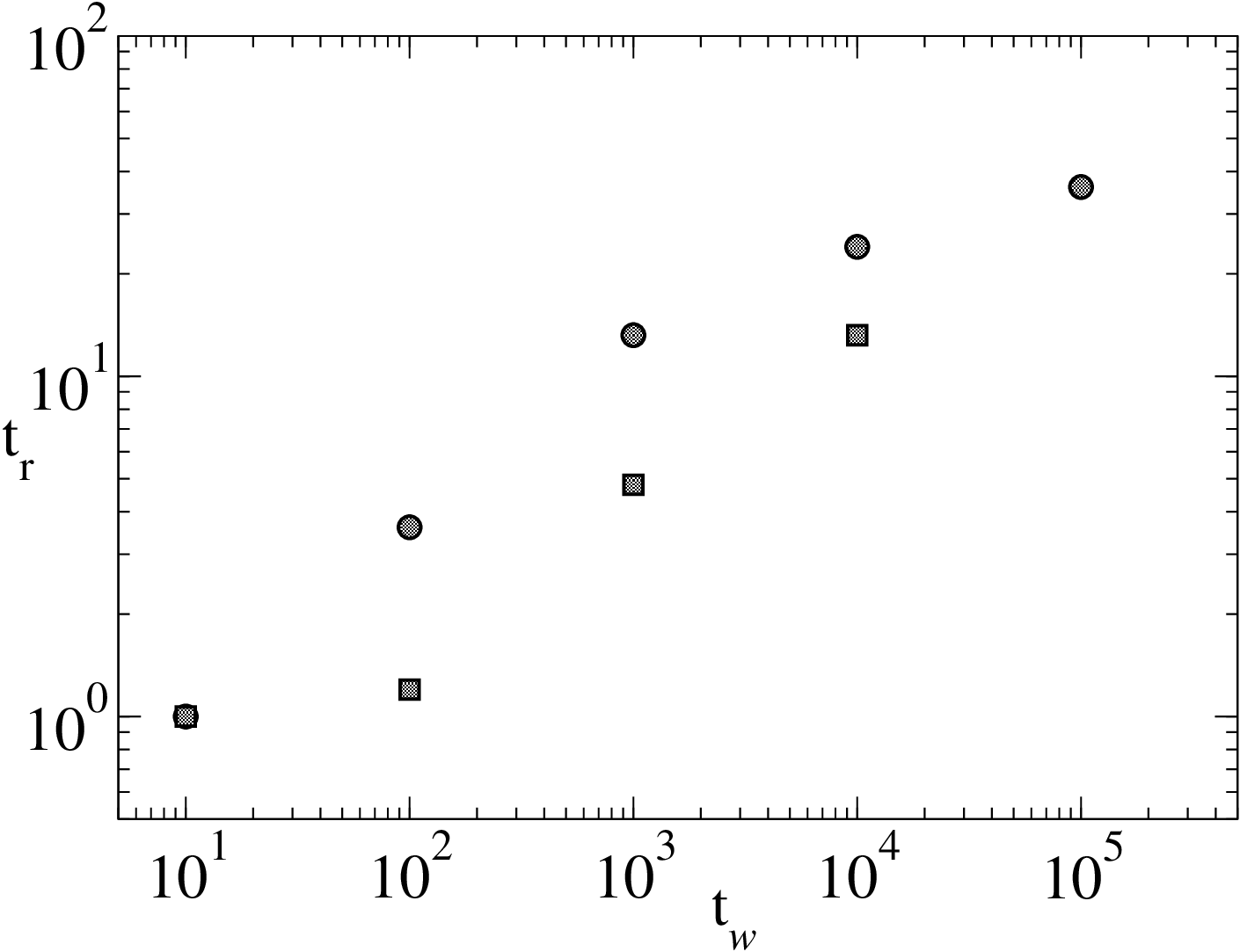}
\caption{\label{fig8} 
	Scaling coefficient used to obtain the scaling in
	Fig.\protect\ref{fig7}a-b. The circle are the scaling
	coefficient for the case in Fig.\protect\ref{fig7}a, the
	square for \protect\ref{fig7}b. }
\end{figure}

A very interesting feature of the correlation functions in the aging
regime is the property of scaling with waiting time.  Theoretical
studies suggest that correlators at different waiting time can be
rescaled by the equation:
\begin{equation}
\label{scaling}
\phi_{\alpha\beta}(q,t_w,t-t_w)=\phi^{st}_{\alpha\beta}(q,t)+
\phi_{\alpha\beta}(q,\frac{h(t-t_w)}{h(t_w)})
\end{equation}
where $\phi^{st}_{\alpha\beta}(q,t)$ accounts for the short time
\cite{Cugliandolo2002}.  The $h(t)=t$ case is the so-called simple aging case.
For a binary mixture of Lennard-Jones spheres the system follows
(\ref{scaling}) with $h(t)=t^\alpha$ and $\alpha\approx 0.88$
\cite{Kob1997b}.  \\ We tried to apply this scaling relation to our
system. We focus on the low temperature quench, since the high
temperature one shows a phenomenology, which is similar to the
Lennard-Jones case. In the case of aging of $T_f=0.3$ no clear scaling
of the correlation functions can be performed. In Fig ~\ref{fig7}a we
show the crude tentative of data scaling. Correlators have been
shifted in time to superimpose the correlators in the region where
$0.7<\phi_q<0.9$. Fig. ~\ref{fig7}b shows the same data, this time
shifted to maximize superposition for $0.35<\phi_q<0.5$. In
Fig.~\ref{fig8} the scaling coefficients $t_r$ used to obtain the
scaling in Fig.~\ref{fig7} are shown as function of the waiting time
$t_w$. It can be seen that even if they could be fitted with power
laws (with an exponent $\alpha\sim0.38\pm0.1$ different from the
exponent found for the Lenard-Jones case), the collapse of the
different $t_w$ curves is not good.


\begin{figure}
\includegraphics*[width=\figwidth]{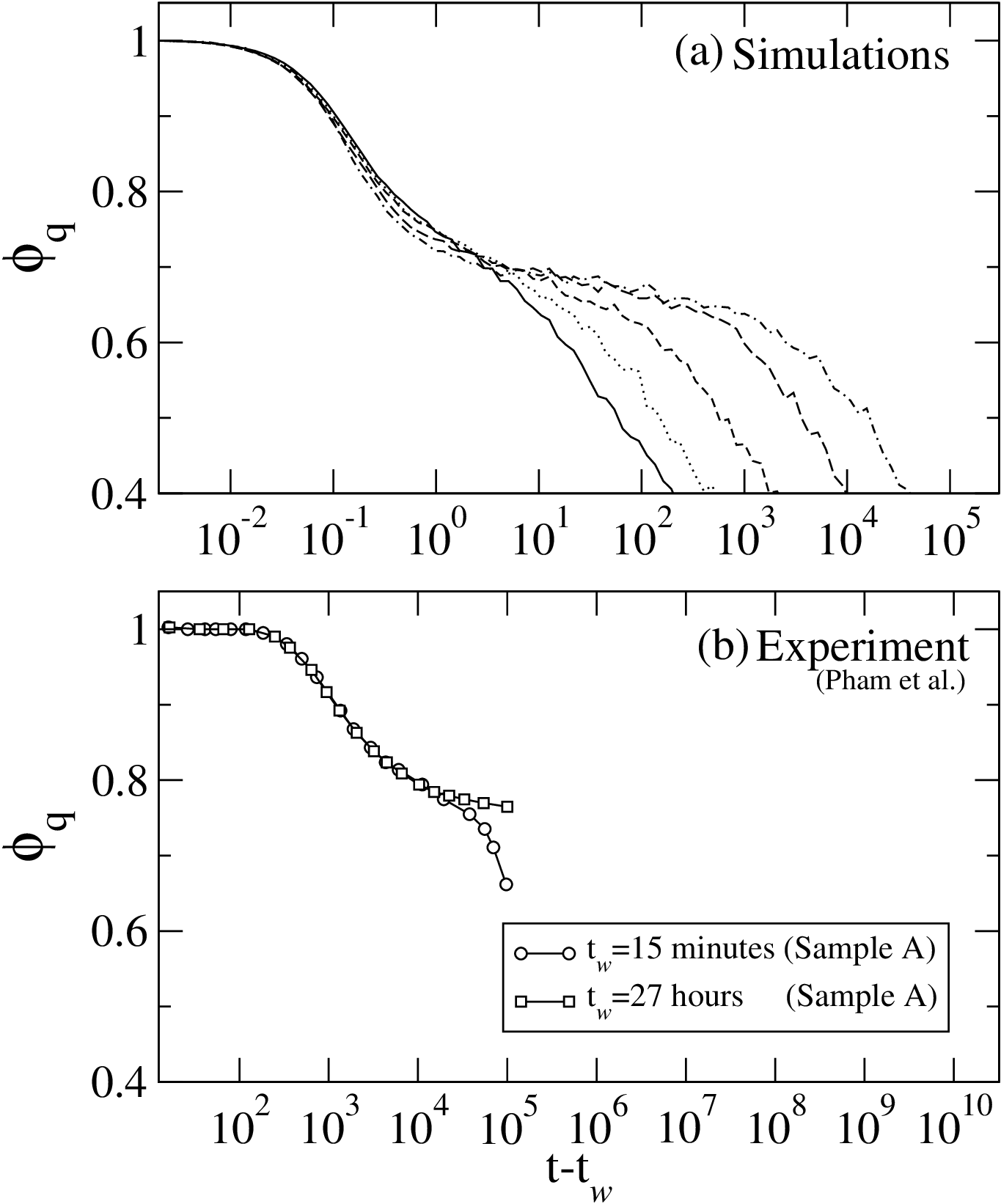}
\caption{\label{fig9} 
      Upper panel: total density-density correlation functions for
      $qR_A=2.93$ at different waiting times for $T_f=1.2$. From right
      to left, longer and longer relaxation time,
      $t_w=10^1,10^2,10^3,10^4,10^5$. \\ Lower Panel: Experimental
      data reproduced from \protect \cite{Pham2003pre}. The unit of
      time are in seconds but time has been rescaled by the
      relative viscosity of the solvent and the wave vector.}
\end{figure}

\begin{figure}
\includegraphics*[width=\figwidth]{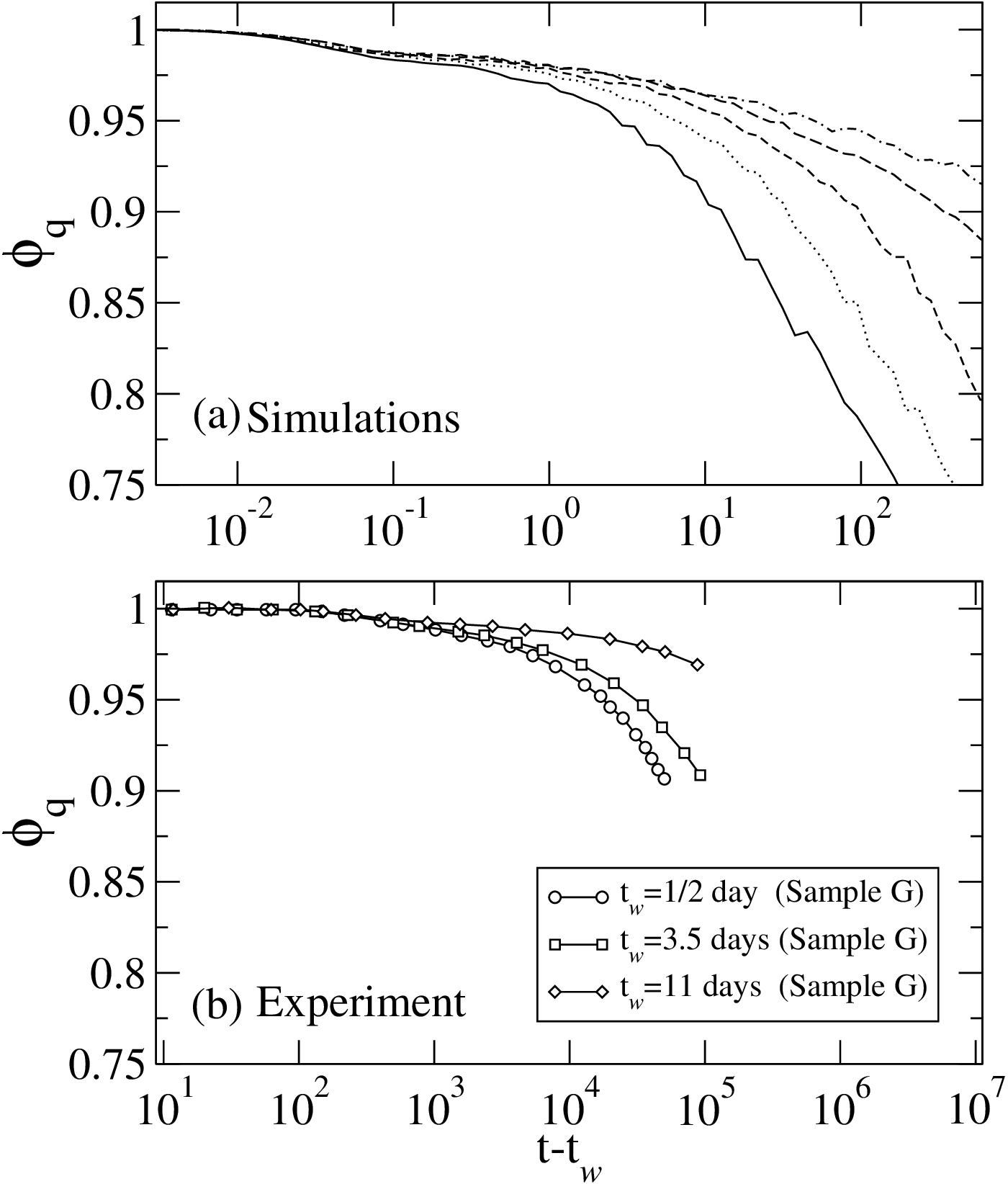}
\caption{\label{fig10} 
      Upper panel: Same as in Fig.\protect\ref{fig9} for $T_f=0.3$.\\ 
      Lower panel: Again 
      experimental data are taken from \protect \cite{Pham2003pre}.}
\end{figure}

It is worth to make a comparison of our numerical results with recent
data measured by Pham et.{\it al} \cite{Pham2003pre}, for a system of
colloids with depletion interaction.  In their experiment, they
measure the density-density correlation function in the aging regime
for two system at different polymer concentration, corresponding
respectively to conditions where the repulsive and the attractive
glass are expected.  To better compare with Pham et. {\it al} data, we
calculated the density dynamical structure factor (irrespective of the
particle type) for the same $q$-vector of the cited
experiment. Results are shown in Fig.~\ref{fig9} and Fig.~\ref{fig10}
where the $T_f=1.2$ and $T_f=0.3$ quenches are respectively plotted
together with the experimental results from Ref. \cite{Pham2003pre}
(lower panel).  Slight differences in the plateau values can be
ascribed to differences in the actual composition of the simulated
(binary mixture) and experimental (slightly polydisperse system). The
agreement between the numerical and the experimental data is
impressive.

Another important two-time observable is the mean squared displacement
(MSD). In  aging, it is defined as $\langle |{\bf
r}(t-t_w)-{\bf r}(t_w)|^2\rangle$, where ${\bf r}(t)$ is the position
of the particle at time $t$ and the average $\langle \cdot
\rangle$ is performed over all the particles and the independent
configurations. Here we  focus on the total MSD, i.e. evaluated with no
distinction between particles of the two species.

\begin{figure}
\includegraphics*[width=\figwidth]{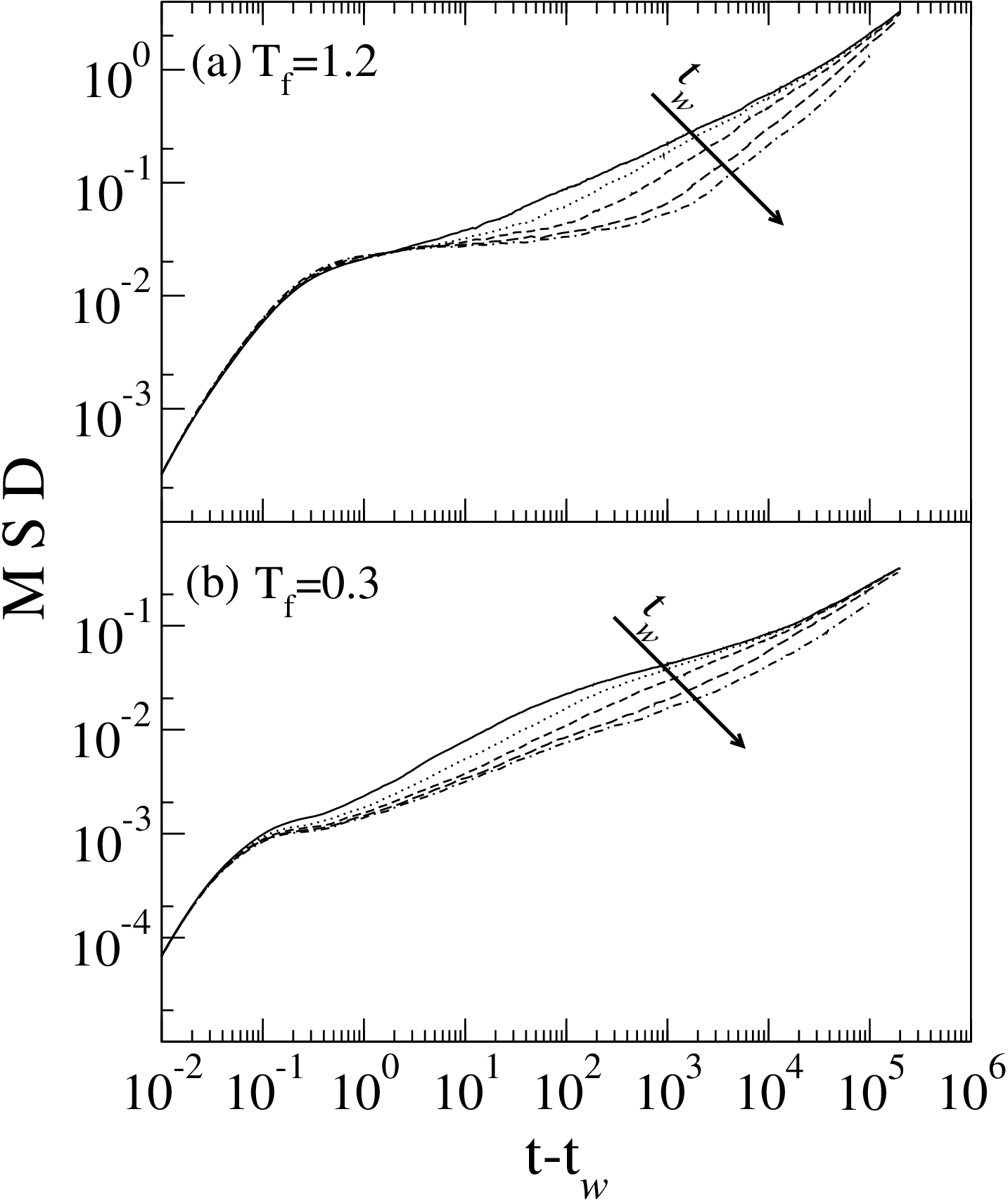}
\caption{\label{fig11} 
  	(a) Mean square displacement (MSD) for the quench at $T_f=1.2$ for
  	increasing waiting times $t_w=10^1$ (continuous line), $10^2$
  	(dotted line), $10^3$ (short dashed line), $10^4$ (long dashed
  	line), $10^5$ (dot dashed line)\\ (b) As in (a) for $T_f=0.3$}
\end{figure}

Fig.~\ref{fig11}a shows the MSD for $T_f=1.2$. In analogy with the
case of the density correlators, at short times there is no
significant $t_w$-dependence. At larger times the diffusive process is
visible.  When the waiting time becomes large, a plateau starts to
emerge at intermediate times, between the ballistic (short time) and
the diffusive (long time) regime. The duration of the plateau grows as
$t_w$ increases. As above, this behavior is similar to the one
observed in previous simulations of a binary mixture of Lenard-Jones
particles \cite{Kob1997b}.

This three time-region behavior (ballistic, plateau, diffusive) is
generally explained in term of the so called cage effect. When
repulsive interactions start to be dominant the system slows down
since each particle starts to get trapped in a shell of first neighbor
particles. The length of the plateau indicates the typical time that a
particle needs to escape this cage. The value of the plateau stands at
the typical size of a cage, i.e. around $10\%$ of the diameter in
agreement with Lindemann's melting criterion \cite{Ashcroft1976}. For
$T_f=0.3$, the MSD shows a completely different behaviour,
(Fig.\ref{fig11}b).  Its evolution can again be divided in three
time-regions, but now the short time (ballistic) and the long time
(diffusive) regions bracket a long intermediate region where the MSD
appears to grow as a power-law in time, with an effective exponent
which decreases with $t_w$. Differently from the previous case, no
evident sign of a plateau is present, except for a hint of a plateau
at MSD$\,=0.001$ which starts to develop for large waiting times for
$t-t_w \approx 0.5$.  This value is consistent with a displacement of
the order of the width of the potential well, and can be interpreted
as a result of particles bonding in the attractive well, i.e. as some
sort of attractive cages.  However, activated bond-breaking processes
appear to destroy such a confinement on time scales longer than
$t-t_w\approx 1$.

\section{Conclusions}
In systems with short-ranged attractive potentials, an efficient
competition between attraction and excluded volume generates a highly
non trivial equilibrium dynamics and two kinetically distinct
glasses. In the ``attractive glass'', the particle localization is
controlled by the bonding distance, while in the``repulsive glass''
localization is controlled by the neighbor location.  This offers the
unique possibility of studying two different aging dynamics within the
same model. In this article, we have realized this numerically for a
model system which shows both types of disordered arrested structures,
the square well potential.  We have found that the aging dynamics is
clearly different for the two glasses, both in the waiting time evolution of
the one-time quantities, and in the time evolution of the two-times
quantities.

The results reported in this article refer to a simple model system
which capture the essence of the physics introduced by the short-range
attraction. The square well potential is indeed a satisfactory
representation of the coarse-grained particle-particle potential in
colloidal system, for example in the presence of depletion
interactions. Of course, dynamical features of the solvent
constituents and of the solvent-particle interactions may
significantly affect the particle dynamics. In this respect, the
experimental validation of the equilibrium results is very valuable.

In particular, we have shown that the partial static structure factor
evolves with $t_w$, showing opposite trends in the two cases, i.e. the
first peak grows with $t_w$ for the repulsive glass while decreases for
the attractive one. We have also shown that the decay of correlation
functions becomes slower and slower on increasing $t_w$ in both cases,
but with noteworthy differences in the strength of the relaxation
(much more intense for the attractive glass case) and in the shape of
the relaxation itself.  The analysis of the MSD brings clear evidence
that, while in the repulsive case there is the emergence of a plateau,
whose time duration increases significantly with $t_w$, in the
attractive glass case this does not happen.  Indeed, in short-ranged
attractive colloids, activated processes can be associated with
thermal fluctuations of order $u_0$, which allow particles to escape
from the bonds.  These processes generate a finite bond lifetime and,
at the same time, may destabilize the attractive glass.  It is
important to find out how the difference in the aging dynamics arising
from the different localization mechanisms are affected by the
presence of activated processes.

Finally, we have shown that a good agreement is found between
experimental and numerical data, also in the aging dynamic.

\begin{acknowledgments}
We acknowledge support from MIUR COFIN 2002 and FIRB and from INFM
Iniziativa Calcolo Parallelo. We thank W.C.K. Poon and K.N. Pham for
useful discussion and for brining their results to our attention.
\end{acknowledgments}

\bibliography{mct,add}

\begin{thebibliography}{38}
\bibitem[{NOT()}]{NOTA}
\bibinfo{note}{Infact the repulsive glass should be named more correctly
  ``hard-sphere'' glass, since this is the real nature of the repulsion in the
  system we shall study.}
\end{thebibliography}
\bibliographystyle{apsrev}

\end{document}